\documentclass[%
 print,superscriptaddress,
onecolumn,
preprintnumbers,
 amsmath,amssymb,
 aps,
floatfix,
]{revtex4-2}
\usepackage{siunitx}
\usepackage{graphicx}
\graphicspath{{./figures/}}
\usepackage{dcolumn}
\usepackage{bm}
\usepackage{hyperref}
\usepackage{soul,xcolor}
\usepackage{caption}
\usepackage{subcaption}
\sethlcolor{red}
\newcommand\hlr[1]{%
  \bgroup
  \hskip0pt\color{white!80!yellow}\hl{%
  #1%
  }
  \egroup
}

\usepackage[english]{babel}

\makeatletter
\def\bbl@set@language#1{%
  \edef\languagename{%
    \ifnum\escapechar=\expandafter`\string#1\@empty
    \else\string#1\@empty\fi}%
  \@ifundefined{babel@language@alias@\languagename}{}{%
    \edef\languagename{\@nameuse{babel@language@alias@\languagename}}%
  }%
  \select@language{\languagename}%
  \expandafter\ifx\csname date\languagename\endcsname\relax\else
    \if@filesw
      \protected@write\@auxout{}{\string\select@language{\languagename}}%
      \bbl@for\bbl@tempa\BabelContentsFiles{%
        \addtocontents{\bbl@tempa}{\xstring\select@language{\languagename}}}%
      \bbl@usehooks{write}{}%
    \fi
  \fi}
\newcommand{\DeclareLangAlias}[2]{%
  \global\@namedef{babel@language@alias@#1}{#2}%
}
\makeatother

\DeclareLangAlias{en}{english}
\DeclareLangAlias{EN}{english}

\begin{document}
\preprint{}
\title{Interfacing spiking VCSEL-neurons with silicon photonics weight banks towards integrated neuromorphic photonic systems}

\author{Matěj Hejda}
\affiliation{%
Institute of Photonics, SUPA Dept of Physics, University of Strathclyde, Glasgow, U.K.
}%

\author{Eli A. Doris}
\affiliation{%
Dept of Electrical and Computer Engineering, Princeton University, Princeton, USA
}%

\author{Simon Bilodeau}
\affiliation{%
Dept of Electrical and Computer Engineering, Princeton University, Princeton, USA
}%

\author{Joshua Robertson}
\affiliation{%
Institute of Photonics, SUPA Dept of Physics, University of Strathclyde, Glasgow, U.K.
}%

\author{Dafydd Owen-Newns}
\affiliation{%
Institute of Photonics, SUPA Dept of Physics, University of Strathclyde, Glasgow, U.K.
}%

\author{Bhavin J. Shastri}
\affiliation{%
Department of Physics, Engineering Physics and Astronomy, Queen’s University, Kingston, Ontario, Canada
}%

\author{Paul R. Prucnal}
\affiliation{%
Dept of Electrical and Computer Engineering, Princeton University, Princeton, USA
}%

\author{Antonio Hurtado}
\email{antonio.hurtado@strath.ac.uk}
\affiliation{%
Institute of Photonics, SUPA Dept of Physics, University of Strathclyde, Glasgow, U.K.
}%

\date{\today}

\begin{abstract}
Spiking neurons and neural networks constitute a fundamental building block for brain-inspired computing, which is posed to benefit significantly from photonic hardware implementations. In this work, we experimentally investigate an interconnected system based on an ultrafast spiking VCSEL-neuron and a silicon photonics (SiPh) integrated micro-ring resonator (MRR) weight bank, and demonstrate two different functional arrangements of these devices. First, we show that MRR weightbanks can be used in conjuction with the spiking VCSEL-neurons to perform amplitude weighting of sub-ns optical spiking signals. Second, we show that a continuous firing VCSEL-neuron can be directly modulated using a locking signal propagated through a single weighting micro-ring, and we utilize this functionality to perform optical spike firing rate-coding via thermal tuning of the micro-ring resonator. Given the significant track record of both integrated weight banks and photonic VCSEL-neurons, we believe these results demonstrate the viability of combining these two classes of devices for use in functional neuromorphic photonic systems.
\end{abstract}

\maketitle


\section{Introduction}\label{sec:intro}
There is an ever-increasing demand for processing of large, complex datasets that is fueled by the exponentially growing amount of available and produced data. Arguably the most advanced data processing capabilities are offered by machine learning (ML) and artificial intelligence (AI) algorithms, driving a massive expansion in scale and complexity of these approaches. While impressive achievements have been demonstrated with large-scale, state-of-the-art AI models, the significant energy requirements and carbon footprint \cite{Strubell2019_X} of these computing approaches cannot be neglected. Furthermore, current general purpose digital chip architectures are starting to approach physical limitations in further downscalling, further strengthening the case for application-specific, optimized computing hardware. By drawing inspiration from the architecture of the brain and its remarkably low power consumption, the field of neuromorphic engineering aims at designing new types of computing architectures that borrow concepts from the operation and structure of biological networks of neurons to increase speed and efficiency of AI computation. There are various technologies actively investigated for neuromorphic computing, with promising candidates including spintronics \cite{Grollier2020_NatElectron}, memristors \cite{Markovic2020_NRP,John2022_NatComm} and photonics \cite{Shastri2021_NatPhot}, among others. 

Spiking neurons and spiking neural networks (SNNs) represent a third generation of artificial neural networks (ANNs) \cite{Maass1997_NeuralNetw} and are of particular interest for future neuromorphic computing platforms. In SNNs, communication between nodes is realized in an sparsely coded, event-based fashion and unlike in previous generations of ANNs, it natively utilizes time for information representation. 
Notably, using sparsely coded spiking signals in the photonic domain provides some desirable additional benefits such as low-loss waveguiding, unparalleled bandwidth of photonic systems and potential for very low energy operation \cite{Miller2017_JLT}. The field of neuromorphic photonics is expanding rapidly, and a wide variety of approaches and photonic devices have been shown to exhibit the desired behaviour and functionalities necessary for such systems. Optical spiking and excitability with direct applications to spike-based information processing have been demonstrated in phase-change materials \cite{Feldmann2019_Nature}, micro-ring resonators \cite{Han2022_Photonics}, photonic crystals \cite{Laporte2018_OE}, superconducting Josephson junction optoelectronics \cite{Shainline2017_PRA}, semiconductor \cite{Nahmias2015_OE} and graphene-based lasers \cite{Jha2022_JLT}, lasers coupled to excitable resonant tunnelling diodes \cite{Ortega-Piwonka2022_OME,Hejda2022_PRAppl}, optical modulators \cite{Mourgias-Alexandris2020_JLT} and semiconductor optical amplifiers \cite{Shi2020_IJSTQE}. 

In this work, we combine together two promising systems in neuromorphic photonics: a spiking artificial photonic neuron based upon a vertical cavity surface emitting lasers (VCSELs) \cite{Skalli2022_OME} and a micro-ring resonator (MRR) weight bank integrated in a silicon photonics platform \cite{Tait2014_JLT, Tait2017_SciRep}. Thanks to their mature fabrication technology that allows for large-scale production and compact device dimensions, VCSELs are already utilized in various practical applications, from telecommunications systems to biometric scanners. VCSELs have also been demonstrated to exhibit a wide range of laser dynamics, with excitable spiking being particularly interesting for the field of neuromorphic engineering. Sub-nanosecond optical spiking regimes can be readily activated \cite{Hurtado2010_OEO_ON} and inhibited \cite{Robertson2017_OL} in injection-locked or electrically-modulated \cite{Robertson2019_IJSTQE} VCSELs to achieve neuron-like spiking dynamics. On the other hand, microring resonators (MRRs) in silicon \cite{Bogaerts2012_LPR} form the basis for many useful photonic integrated circuits (PICs), including integrated true time delay \cite{Xiang2018_IJSTQE}, beam steering \cite{Larocque2019_OEO}, dense wavelength-division multiplexing (DWDM) for high-performance computing \cite{Liang2020_OFCCO22PT}, neurosynaptic networks with self-learning capabilities \cite{Feldmann2019_Nature}, machine learning \cite{deLima2019_JLT,Huang2021_NatElectron}, and more. MRR-based weight banks in particular \cite{Tait2014_JLT,Tait2017_SciRep} have found success as the basis for the linear front end of photonic neurons (i.e., photonic synapses) in machine learning applications \cite{deLima2019_JLT,Huang2021_NatElectron}. Since MRR weight banks are inherently scalable via wavelength-division multiplexing (WDM), demonstrating their use in the context of spiking neural networks indicates great promise for expanding to the more highly-interconnected network architectures that are necessary for more complex tasks.

This paper provides an experimental demonstration of two functional layouts using VCSELs together with SiPh MRRs: synaptic spike amplitude weighting of the high-speed (sub-nanosecond long) VCSEL-neuron spiking signals using a single micro-ring in the weight bank (demonstrating functionality theoretically proposed in \cite{Tait2014_JLT}), and a spike rate coding mechanism using a continuously (tonic) firing VCSEL-neuron directly modulated with locking (injection) signal provided by the micro-ring. These two different experimental layouts demonstrate how these two systems can cooperate in various ways, achieving useful functionality for spike-based information processing. The manuscript is divided as follows: Section \ref{sec:characterization} introduces the two devices used in the experiments and provides their characteristics. In Section \ref{sec:VCSELtoMRR}, we describe the experimental setup and achieved results for VCSEL-neuron spike weighting in the micro-ring and in Section \ref{sec:MRRtoVCSEL}, we demonstrate spiking rate coding in the VCSEL-neuron with injection power modulation via the weight bank micro-ring resonator. Finally, discussion and conclusions are provided in Section 
\ref{sec:conclusions}.
\begin{figure*}[ht!]
     \centering
     \includegraphics[width=0.95\textwidth]{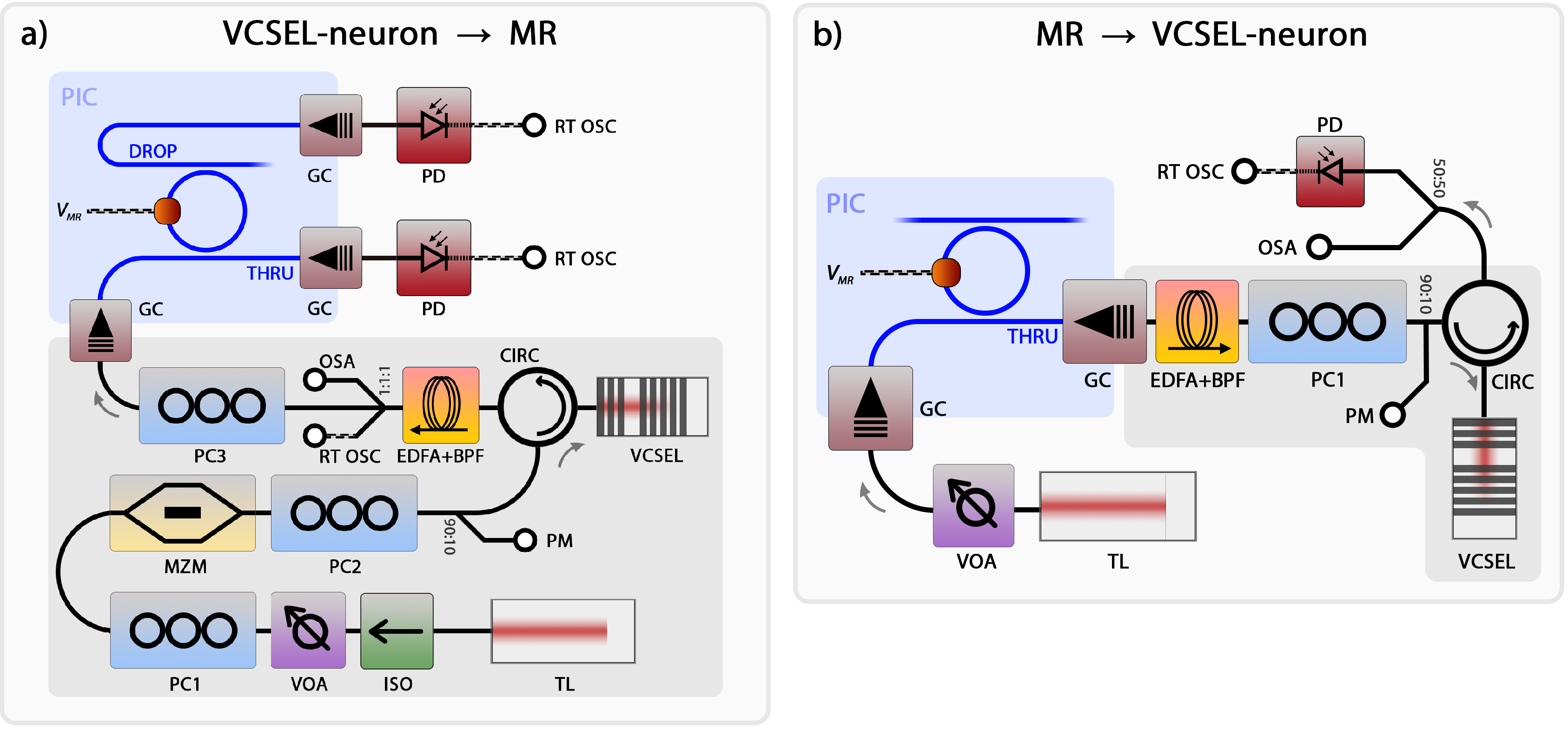}
     \caption{Experimental setups for a) spike weighting in the micro-ring, and b) direct VCSEL-modulation via signal coming through a micro-ring. TL - tuneable laser; ISO - isolator; VOA - variable optical attenuator; PC(1,2,3) - polarization controllers; MZM - Mach-Zehnder modulators; PM - power meter; CIRC - circulator; VCSEL - vertical cavity surface emitting laser; EDFA - erbium-doped fiber amplifier; BPF - bandpass filter; RT OSC - real-time oscilloscope; OSA - optical spectrum analyzer; GC - grating coupler; PD - photodetector. Ratios indicdated are for the power couplers/splitters.} 
     \label{fig:ExpSetups}
\end{figure*}
\section{Devices characterization}\label{sec:characterization}
In this work, we use a telecom-wavelength VCSELs operating at the standard telecom wavelength of \SI{1550}{\nano\metre}. The emission spectrum of the VCSEL exhibits two peaks corresponding to the two orthogonal polarizations of the device’s fundamental transverse mode. The higher wavelength mode is dominant (dominant lasing mode) while the other remains suppressed (subsidiary mode). Using notation from previous works \cite{Robertson2020_IJSTQE}, these two modes can also be referred to as the orthogonal and parallel polarizations of the fundamental mode. These two spectral peaks are separated by approximately \SI{250}{\pico\metre}. The lasing threshold is approx. $I_{T} \approx $ \SI{1.5}{\milli\ampere}. VCSELs are known to exhibit a wide range of nonlinear dynamical behaviors. To achieve excitability and spiking dynamics, the laser is injection locked to a polarization-matched external signal with a small negative frequency detuning $\Delta_f$. In such an operational regime, amplitude (power) perturbations of the injection signal may elicit sub-ns spiking responses if the perturbation is sufficiently strong (crossing a "threshold" power value). The spiking responses from the VCSEL-neuron exhibit all the typical hallmarks of excitability such as all-or-nothing character and refractory (lethargic) period with dynamical behavior analogous to leaky integrate-and-fire neuronal models \cite{Robertson2020_SciRep}.
\begin{figure*}[t]
     \centering
     \includegraphics[width=0.85\textwidth]{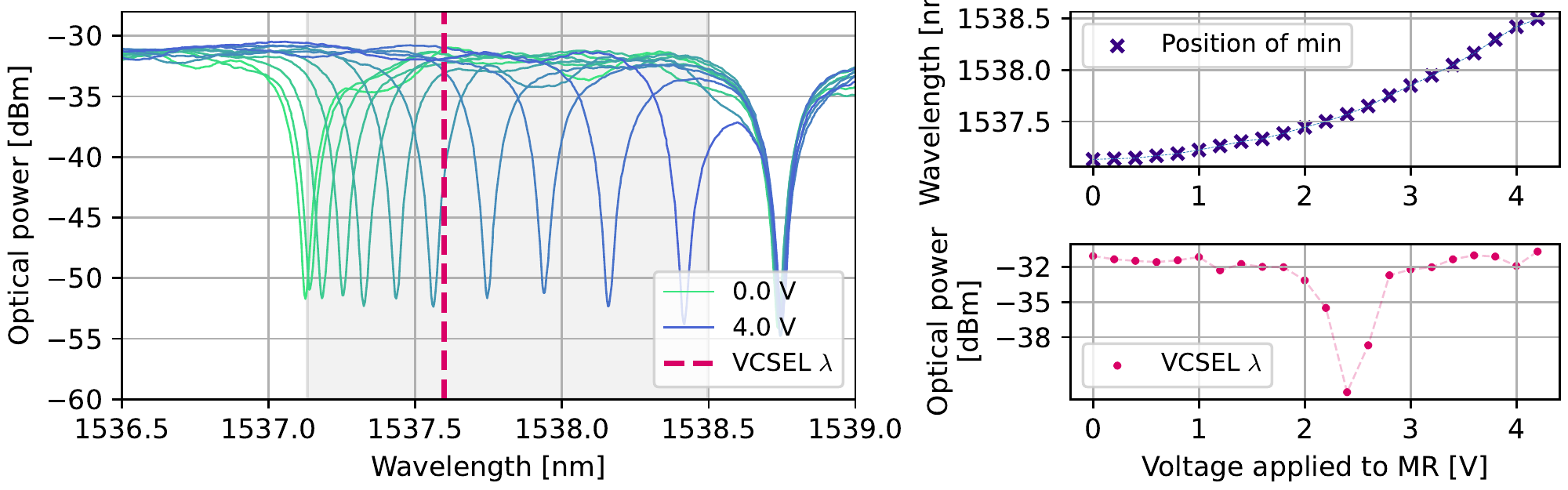}     
     \includegraphics[width=0.85\textwidth]{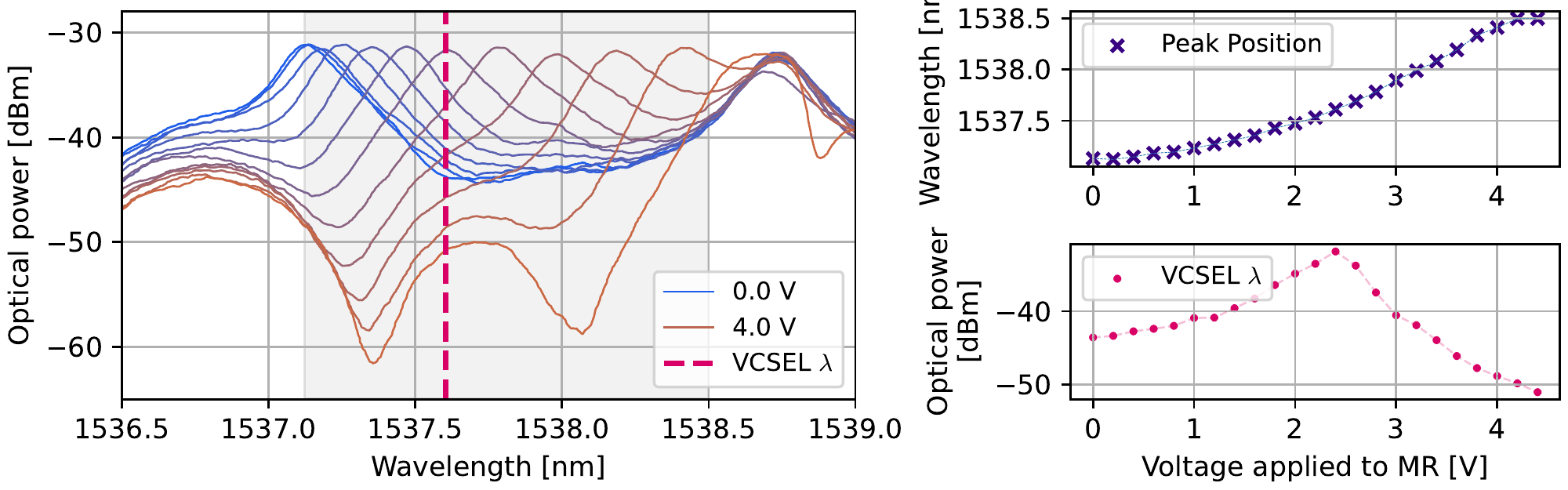}
     \caption{micro-ring characterization: THRU port readout (top) and DROP port readout (bottom), with power at fixed wavelength (dashed line) and maxima (minima) position as function of $V_{\mathrm{MR}}$.} 
     \label{fig:MRR_spectra}
\end{figure*}

The micro-ring resonator (MRR) used for this experiment forms part of a 4 MRR weight bank \cite{Tait2017_SciRep} that allows for broadcast and weight operation \cite{Tait2014_JLT}. In this weight bank, all MRRs are in the "add-drop" configuration with adjacent THRU/IN ports connected and adjacent ADD/DROP ports connected. This weight bank is on a chip fabricated by Advanced Micro Foundry (AMF) in a multi-project wafer (MPW) run coordinated by CMC Microsystems organization. The core layer is 220 nm thick Si, with 500 nm wide bus and ring waveguides. Bus-coupler gaps are 200 nm, and the four MRRs have radii around 8 $\mu$m varied slightly to produce approximately 1 nm resonance spacing. The DROP and THRU port spectra of this weight bank are shown in Fig. \ref{fig:MRR_spectra}, highlighting an actuated resonance (gray region) along with a nearby fixed resonance. As expected, the lack of change in resonance of the non-actuated MRR as the actuated MRR is changed indicates low crosstalk and independent MRR operation. Also shown are the expected power at a given fixed wavelength (highlighted by a dashed red line) and position of extrema of the actuated as a function of the bias voltage applied to the heater element controlling the resonant frequency of the MRR.

\section{Post-neuronal weighting: VCSEL $\rightarrow$ MRR}\label{sec:VCSELtoMRR}

First, we investigate the viability of an integrated MRR as part of a photonic weight bank to controllably weight the amplitude of optical spiking signals delivered by a VCSEL-neuron. The sub-ns spikes to be weighted were deterministically elicited in the VCSEL when subject to the injection of externally-modulated light signals \cite{Robertson2020_IJSTQE}. Afterwards, the spiking signals were coupled to an integrated photonic circuit, where a single micro-ring was used to alter the signal (spike) amplitude via thermal control of a ring heater element.

\subsection{Methods}
The experimental setup for the VCSEL-MRR system is shown in Fig. \ref{fig:ExpSetups}(a).
The optically injected signals into the VCSEL neuron are provided by a tunable laser (EMCORE micro-ITLA).
The modulation of the optical injection line was generated on an AWG (Tektronix AWG7122B, 12 GSa $s^{-1}$) and realized using a fiber-based Mach-Zehnder modulator and had a form of square pulses of \SI{750}{\pico\second} (8 Sa) with \SI{4.66}{\nano\second} (8$\times$7 Sa) spacing, using p-p DAC range of \SI{600}{\milli\volt} in the AWG. Prior to entering the VCSEL, the injection lane was split with a 90:10 fiber optic splitter, with the 10\% branch used for injection power monitoring. The 90\% power branch is fed into the VCSEL through a circulator, which also enables to obtain the VCSEL's output signal. The VCSEL was biased at \SI{5}{\milli\ampere} (threshold of \SI{1.5}{\milli\ampere}), the injection detuning was approx. \SI{-3}{\giga\hertz} with respect to the orthogonally-polarised mode of the device. and the injection line power was \SI{170}{\micro\watt}. The VCSEL (circulator) output was then amplified using an EDFA (PriTel) and equally split using a 1-to-4 fiber splitter (Thorlabs). These outputs were used to provide the signal into the SiPh chip, as well as to directly monitor the VCSEL output time trace via a photodetector and a real-time oscilloscope (PD, RT OSC; Discovery Semiconductors Lab Buddy, Tektronix MSO/DPO70000) and the VCSEL's optical spectrum on an optical spectrum analyzer (OSA; Apex Technologies).

During this experiment, the injection-locked VCSEL neuron was continuously perturbed with the dedicated sequence of square perturbations modulating the injection, providing deterministic, equidistant single optical sub-ns spikes at the VCSEL output. To perform the weighting, the selected micro-ring on the SiPh chip was swept through with a given DC voltage $V_{\mathrm{MR}}$ on each sweep step. Light from optical fibers was coupled to the chips via foundry-provided grating couplers optimized for 1550 nm light and an incident angle of 8 degrees. An 8-fiber V-groove array polished at 8 degrees, manufactured by OZ Optics, was positioned above an array of 8 grating couplers on chip. The first and last grating couplers are directly connected with a bus waveguide, allowing for optical alignment and loss measurement by monitoring a constant-power alignment signal. An input signal of 10 dBm was injected into the alignment port, and at optimal alignment a power of approximately -3 dBm was mesaured at the output photodetector. This indicates a loss per grating coupler of approximately 6-7 dBm, which is typical for such structures in practical use. 

In the future, proving that VCSEL neurons can be directly cascaded will allow to reduce many sources of loss, as well as it will allow to remove the EDFAs from the optical path. To that end, photonic wire bonding \cite{Lindenmann2012_OEO} presents a potential opportunity for sub-1-dB losses \cite{Blaicher2020_LightSciAppl} while coupling from an optical fiber to a photonic chip. Similarly, lower losses are often possible with edge couplers \cite{Fang2009_IPTL,Pu2010_OC}, although these are typically prone to environment-induced instabilities in coupling efficiency that are not present in photonic wire bonds.

\begin{figure}[t!]
     \centering
     \includegraphics[width=0.6\linewidth]{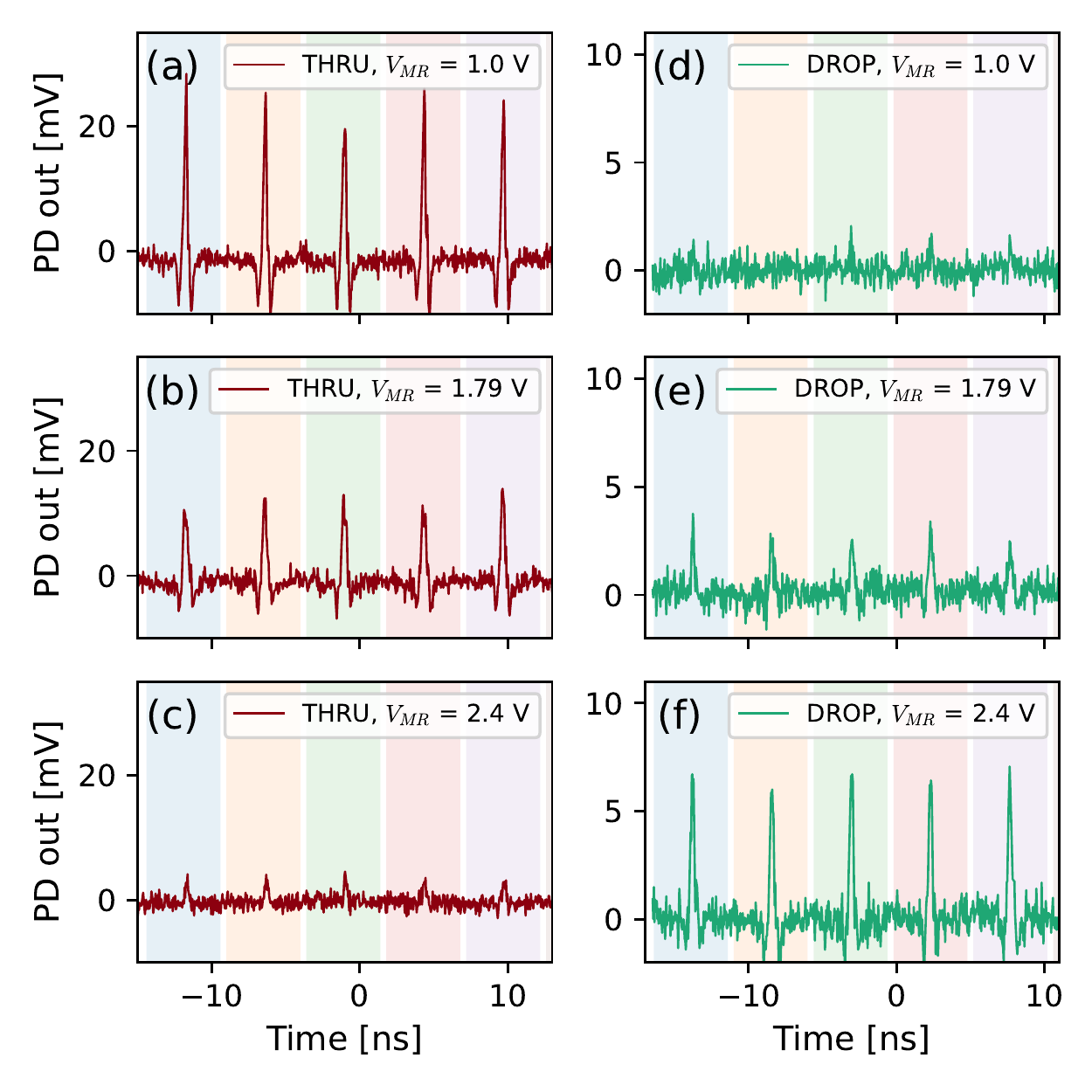}
     \caption{Sequences of deterministically triggered excitable spikes in the VCSEL-neuron after passing through the weighting micro-ring for three different values of $V_{\mathrm{MR}}$. Traces were recorded on both outputs of the micro-ring: the THRU port (left column, in red) and the DROP port (right column, in green).} 
     \label{fig:weighted_traces}
\end{figure}

\begin{figure*}[t!]
     \centering
     \includegraphics[width=0.87\textwidth]{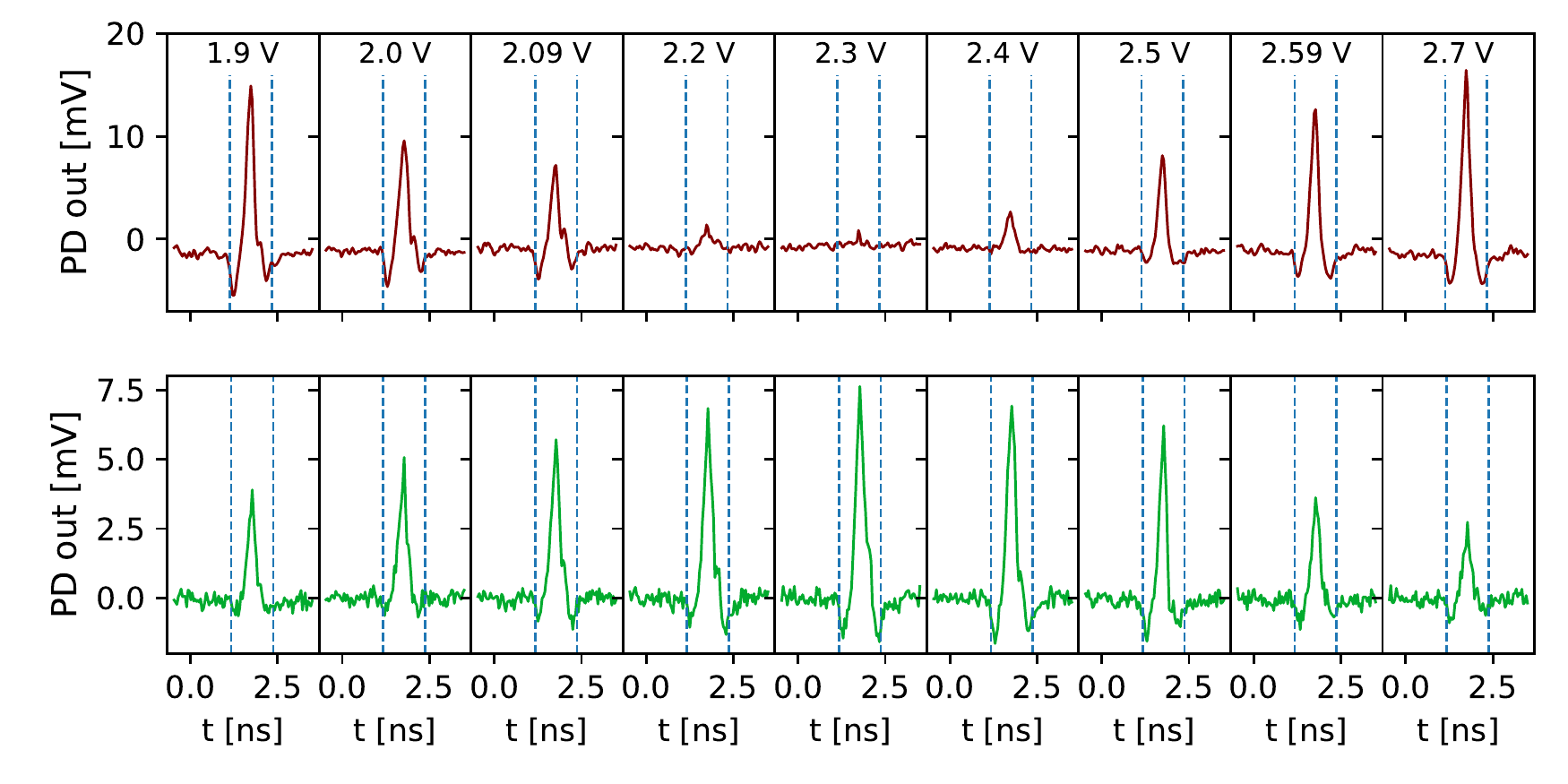}
     \caption{Demonstration of spike amplitude weighting at both THRU (top row) and DROP (bottom row) ports of the micro-ring resonator output. In each case, the spike trace shown corresponds to a mean spike envelope calculated from all the recorded repetitions for a given value of $V_{\mathrm{MR}}$.} 
     \label{fig:weighted_mean_spikes}
\end{figure*}
\subsection{Results}
\begin{figure}[h!]
     \centering
     \includegraphics[width=0.45\textwidth]{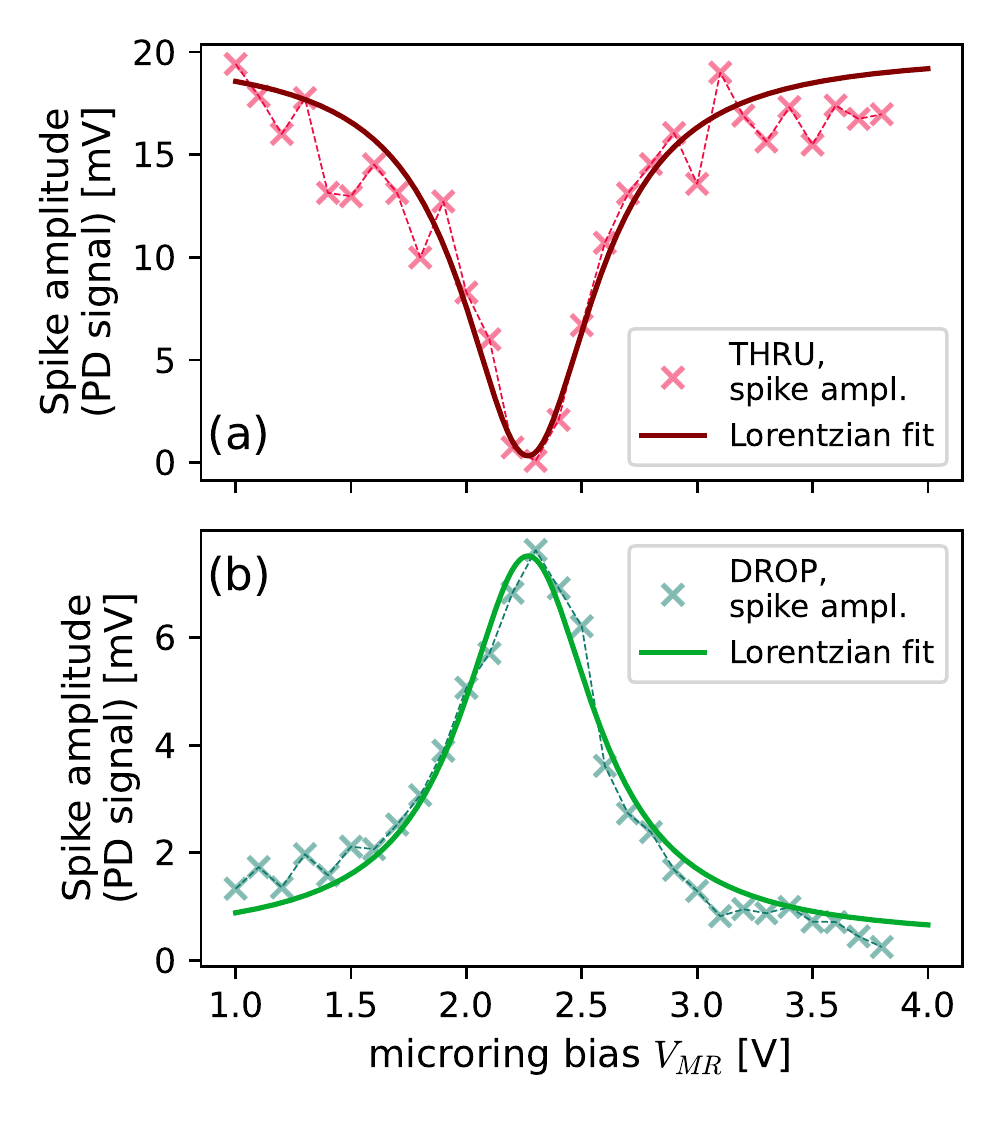}
     \caption{Mean amplitudes of all spikes ($n_s=7$ per measurement, $n_m=15$ repetitions) as a function of micro-ring heater bias $V_{\mathrm{MR}}$ with Lorenzian fit (solid line) using the least squares method. The two plots show (a) THRU; (b) DROP port readouts.} 
     \label{fig:weighting_FoM}
\end{figure}

For each $V_{\mathrm{MR}}$, a time trace with $n_s=7$ spikes was acquired. In total, 29 voltage steps were recorded, from $V_{\mathrm{MR}}$ = \SI{1}{\volt} to \SI{3.9}{\volt} with \SI{100}{\milli\volt} increments. This sweep was repeated $n_m=15$ times for statistical evaluation. In each case, traces were recorded from: 1) the direct VCSEL output, 2) the THRU port of the MRR, and 3) the DROP port of the MRR. All time traces collected from the THRU port of the micro-ring are shown in red, and all the traces collected from the DROP port are printed in green. An example of a single recorded time trace for three selected values of $V_{\mathrm{MR}}$ from both ports is shown in Fig. \ref{fig:weighted_traces}, clearly showing how sweeping the micro-ring resonance alters the recorded spike amplitude. Furthermore, each of these $n_m=15$ traces taken with equivalent experiment parameters was sliced into small temporal bins, including within each a single MRR-weighted spike (these bins are shown as colored shading in Fig. \ref{fig:weighted_traces}). Within each bin, the maxima (spike peak) was found, and using the position of this spike peak as an alignment reference, all the bin traces containing a single spike ($15\times7$ total) were overlaid and averaged over to obtain a representative weighted spike for a given $V_{\mathrm{MR}}$ from all the repeated measurements while avoiding possible averaging issues coming from small trace misalignment or jittering. These mean representative spikes are shown in Fig. \ref{fig:weighted_mean_spikes}. After calculating the representative spike, its maxima is saved as representative spike amplitude. The same procedure was followed for the DROP and THRU ports of the MRR. The resulting representative peak spike amplitudes are shown in Fig. \ref{fig:weighting_FoM}(a,b). Both figures were fitted with a Lorentzian in the interval around the extrema, showing a good degree of agreement in this range of micro-ring bias voltages. The THRU photodetector time traces did not contain the DC signal component (constant lasing emission from the VCSEL operated above the threshold). Therefore, this DC component was also numerically filtered from the DROP port signal to compare signals of equivalent nature.


\section{Pre-neuronal modulation:\\MRR $\rightarrow$ VCSEL}\label{sec:MRRtoVCSEL}
In this Section, we utilize the weighted output from a single micro-ring in the integrated weight bank, and injection lock the VCSEL to this signal to realize a spiking VCSEL-neuron. By directly modulating the constant locking signal via the micro-ring and changing its power, we introduce continuous spiking dynamics in the VCSEL-neuron, where the spike firing frequency is inversely proportional to the input power \cite{Hejda2020_JPP}. The capability of VCSELs to directly encode analog signals into continuous, real-time encoded spike trains that represent input signal amplitude into local spike firing rate \cite{Hejda2020_JPP} holds promise for utilizing VCSELs as spike-domain input encoders. This functionality was previously demonstrated on multichannel (RGB) digital images \cite{Hejda2021_APLPhot} using a time-domain multiplexed, implementation-friendly single VCSEL-neuron layout. 
\begin{figure*}[t!]
    \centering
    \begin{subfigure}[b]{0.44\textwidth}
         \centering
         \includegraphics[width=\textwidth]{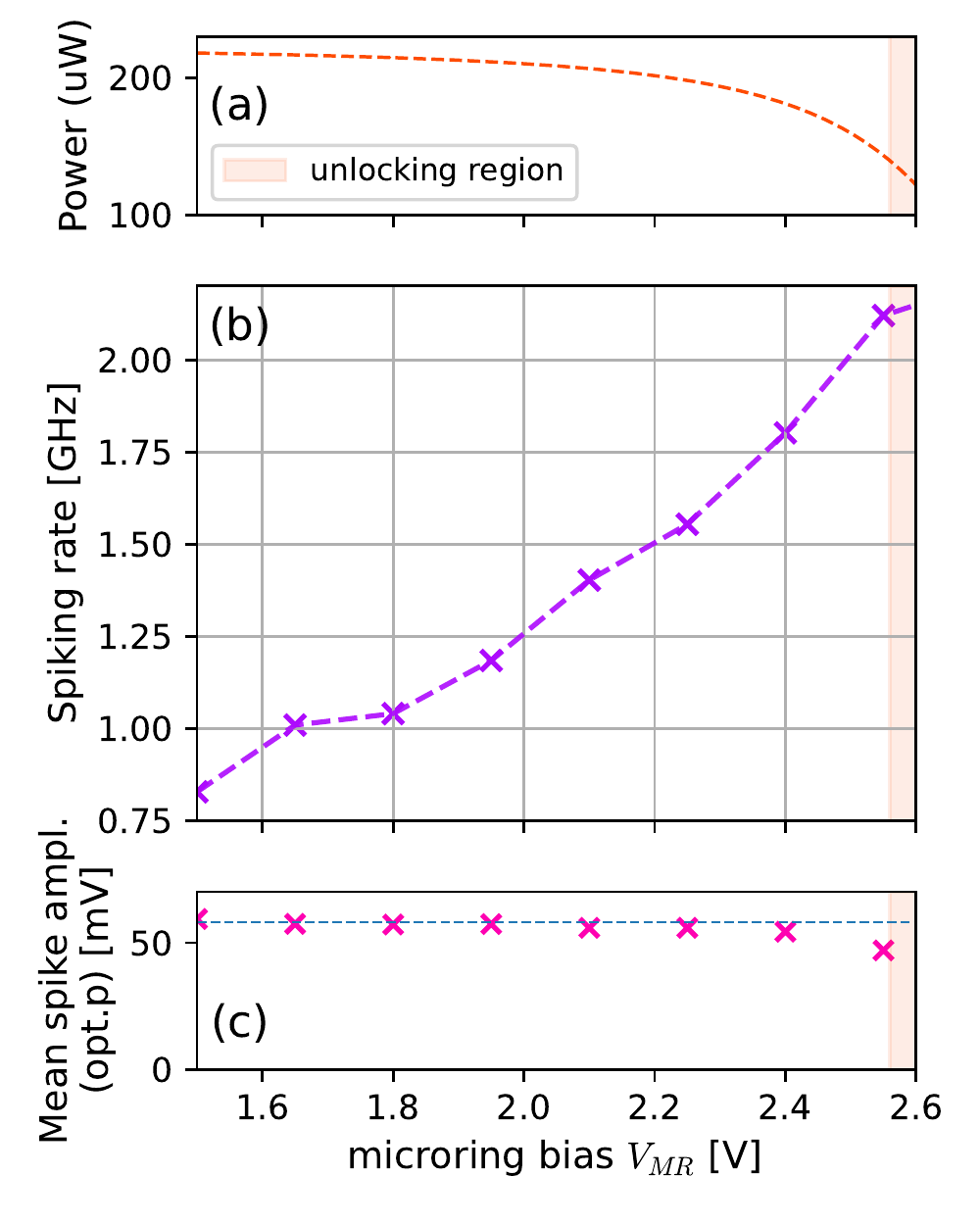}
    \end{subfigure} 
    \begin{subfigure}[b]{0.44\textwidth}
         \centering
         \includegraphics[width=\textwidth]{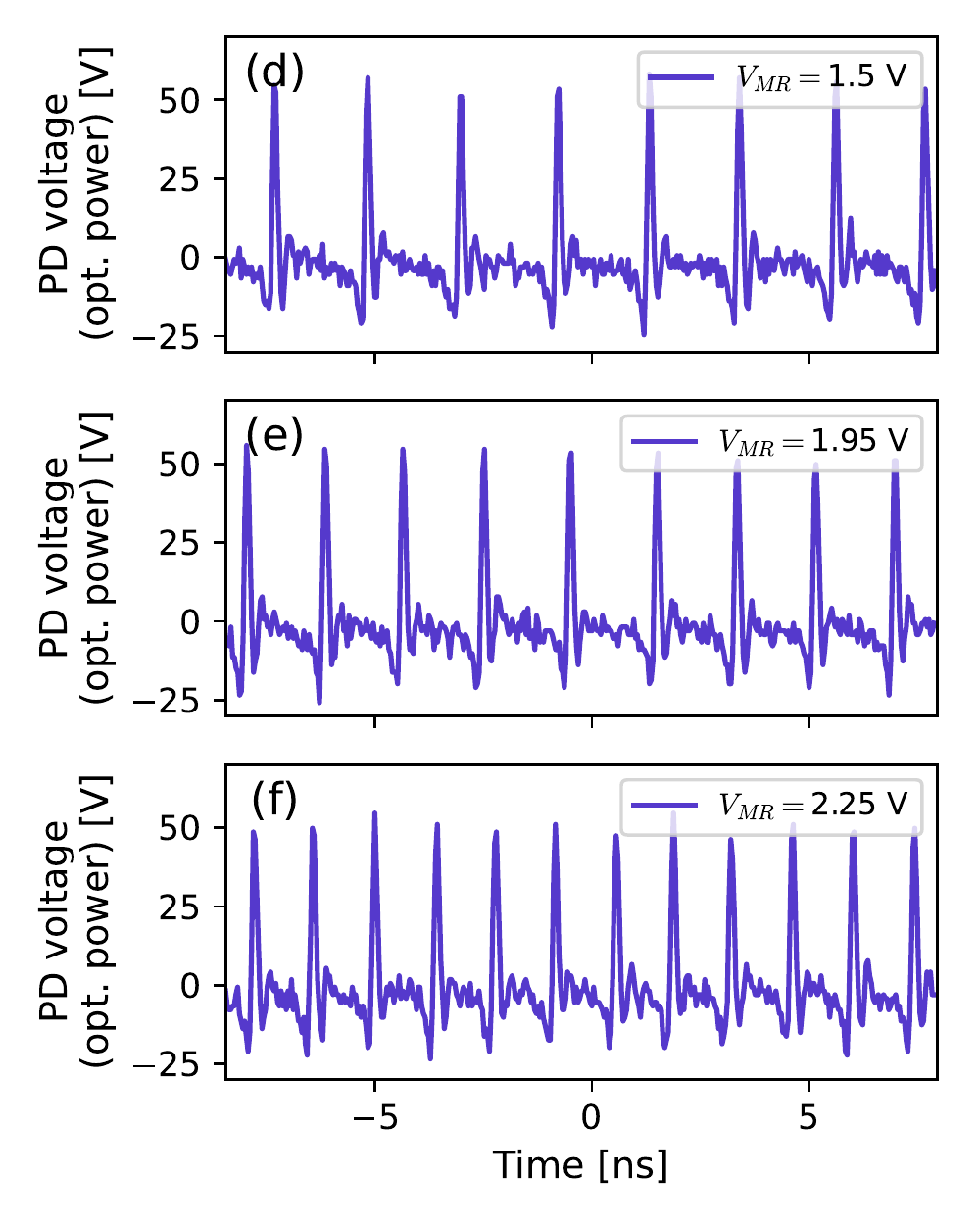}
    \end{subfigure}     
  
    \caption{Demonstration of rate coding in spiking VCSEL-neuron injection locked to a signal coming from the MRR. (a) Injection signal power as a function of $V_{\mathrm{MR}}$. (b) Spike firing rate in the VCSEL-neuron as a function of $V_{\mathrm{MR}}$. (c) Mean spike amplitudes within the spike train. (d-f) Examples of continuous spiking traces produced by the VCSEL-neuron locked to the signal from the MRR. As the voltage applied to the micro-ring is gradually increased from (a) \SI{1.5}{\volt} through (b) \SI{1.95}{\volt} to (c) \SI{2.25}{\volt}, the spike firing rate monotonically increases.} 
    \label{fig:rate_coding}
\end{figure*}
\subsection{Methods}

To realize the direct injection locking of the VCSEL to a signal coming from the SiPh chip, a micro-ring was selected whose resonance is wavelength matched to the VCSEL emission. The signal from the tunable laser (EMCORE micro-ITLA) passed through a variable optical attenuator and was fed through the weighting micro-ring. The THRU port of this ring was then coupled off the chip into a fiber. After the micro-ring-weighted signal was collected from the chip, it was amplified via an EDFA (with bandpass filter) and injected into the VCSEL through a polarization controller and circulator, with injection power $P_{inj}\approx$ \SI{220}{\micro\watt}. This signal was polarization adjusted to ensure locking into the orthogonal (dominant) polarization, and wavelength adjusted to ensure injection detuning of approximately \SI{-3.75}{\giga\hertz}. The output of the VCSEL-neuron was then collected through the circulator and recorded using an amplified photodetector (Discovery Semi LabBuddy) on a real-time oscilloscope.
\subsection{Results}

The MRR → VCSEL experiment results are shown in Fig. \ref{fig:rate_coding}. A sweep of the micro-ring bias voltage $V_{\mathrm{MR}}$ was performed, from \SI{1.5}{\volt} to \SI{4.2}{\volt} with \SI{150}{\milli\volt} increments (total 21 steps). For each $V_{\mathrm{MR}}$, a \SI{40}{\nano\second} trace was recorded, where individual spikes were counted and related to the recording length ($t_{len}$=\SI{40}{\nano\second}) to obtain the local spiking rate. This procedure was repeated $n=12$ times, yielding the representative firing rate (and its dependence on applied micro-ring bias $V_{\mathrm{MR}}$) as a mean of these $n=12$ local firing rates. 

The optical power of the locking signal to the VCSEL as a function of $V_{\mathrm{MR}}$ is shown in Fig. \ref{fig:rate_coding}(a). As the resonance of the micro-ring shifts during the $V_{\mathrm{MR}}$ sweep towards the wavelength of the locking signal, the power present at the output THRU port drops down. In the injection locked layout used for the VCSEL-neuron, this means weakening of the locking signal. For a VCSEL in a continuously firing regime, this increases the spiking frequency \cite{Hejda2020_JPP} followed by a transition to different dynamical phenomena. This is demonstrated in Fig. \ref{fig:rate_coding}(b) by the monotonically increasing spike firing rate. Under the given set of experimental conditions at the start of the experiment (with the micro-ring resonance frequency far from the locking signal frequency), the VCSEL-neuron fires continuously with a mean spiking rate of $\approx$ \SI{750}{\mega\hertz} that increases towards $\approx$ \SI{2.25}{\giga\hertz}, which approximately represents the highest achievable spike firing rate in this VCSEL, in this case achieved for $V_{MR} = $ \SI{2.55}{\volt}. The spike firing rate modulation effect is also demonstrated by examples of VCSEL-neuron output traces for different values of $V_{\mathrm{MR}}$ shown in Fig. \ref{fig:rate_coding}(d-f).

By further increasing the micro-ring bias, the power at the THRU port drops significantly as the ring resonance is approached. In the VCSEL-neuron, for such a significant reduction of input injection power, we have observed a gradual transition from continuous, distinct repeated spikes to period-1 resembling dynamics with a noticeable lowering of oscillation amplitude, followed by breaking of injection locking and the oscillation-like dynamics. This is shown in Fig. \ref{fig:rate_coding}(c) as a sudden decrease in recorded mean spike amplitude. For such low levels of injected signal, the injection locking of the VCSEL-neuron is (temporarily) broken, and the VCSEL behaves as a free-running laser without exhibiting spiking dynamics. Therefore, to realize rate-coding in the VCSEL-neuron using this layout, only part of the ring's dynamical range should be used to avoid unlocking close to the ring resonance. 

\section{Discussion and Conclusions}\label{sec:conclusions}

This work provides the first experimental demonstration of interconnectivity between spiking photonic VCSEL-neurons and integrated MRR weight banks, with a focus on two key relevant functions for future scaled-up photonic neural computing: spike weighting and all-optical rate-encoding. In the current first iteration of these functional layouts, both were utilized using a telecom-wavelength VCSEL coupled off-chip to the PIC with the micro-ring (weight bank) via grating couplers. While coupling of the off-chip laser allows for additional flexibility and control when realizing the experiments, the ultimate goal lies in a system realized fully on-chip. Such a system would provide significantly increased utility, particularly due to the reduction in overall spatial footprint and likely also better control over environment-induced noise.
Furthermore, we believe that the effects of polarization of the VCSEL output during the spike may provide additional interesting functionality in wavelength-sensitive weighting, such as in the micro-ring. As excitable optical spikes are elicited via perturbation in the injection to the VCSEL-neuron, the laser undergoes an excursion that may involve brief switching between the two mutually orthogonal polarizations while the neuron is active \cite{Hurtado2010_OEO_ON}. This may hold potential for more advanced weighting schemes in combined VCSEL-MRR systems, and will be further investigated by the authors.


In conclusion, we have experimentally demonstrated the joint operation of a spiking VCSEL-neuron and a micro-ring resonator weight bank for operation in all-optical spike-based information processing. A spike-weighting layout (VCSEL → MRR) and a direct rate-encoding (MRR → VCSEL) layout have been validated, demonstrating both systems as being mutually compatible and viable building blocks for larger, hybrid photonic spiking networks with sub-ns spike times. This work opens new routes towards novel, more advanced solutions for fully on-chip layouts combining lasers as spiking neurons and integrated weight banks for future functional neuromorphic computing photonic hardware

\section{Acknowledgements}\label{sec:ack}
The authors acknowledge support from the European Commission (Grant 828841-ChipAIH2020-FETOPEN-2018-2020) and by the UK Research and Innovation (UKRI) Turing AI Acceleration Fellowships Program 4-(EP/V025198/1). S. Bilodeau acknowledges funding from the Fonds de recherche du Québec - Nature et technologies. B. Shastri acknowledges funding from the Natural Sciences and Engineering Research Council of Canada (NSERC).

\bibliography{main}

\end{document}